\documentclass[11pt]{article}
\usepackage[dvips]{graphicx}
\usepackage{amsfonts}
\usepackage[english]{babel}
\usepackage[dvips]{color}
\usepackage{amsmath}
\usepackage{amsthm}
\usepackage{amssymb}
\usepackage{amsfonts}
\usepackage{xcolor}
\usepackage{caption}
\usepackage{amsthm}
\usepackage{xcolor}

\def\vsn{\par \noindent}
\def\vvsn{\vskip 0.5truecm\noindent}
\begin{document}
\begin{center}
{\Large {\bf
{A comparative view of Becker, Lomnitz, \\ and Lambert linear
viscoelastic models}}}
\vvsn
\\ {{\bf Juan Luis Gonzales-Santander}$^a$, 
{\bf Francesco Mainardi}$^b$
}
\vvsn
 $^a$
 Orcid:{0000-0001-5348-4967}
 \\ Department  of Mathematics, University of Oviedo,\\
C Leopoldo Calvo Sotelo 18, E-33007 Oviedo, Spain\\
\vsn
$^b$ 
Orcid:{0000-0003-4858-7309}
\\ Department of Physics $\&$ Astronomy,
University of Bologna, and INFN,
\\ Via Irnerio 46, I-40126 Bologna, Italy
\end{center}
\begin{abstract}
We compare the classical viscoelastic models due to Becker and Lomnitz with
respect to a recent viscoelastic model based on the Lambert $W$ function. We
take advantage of this comparison to derive new analytical expressions for
the relaxation spectrum in the Becker and Lomnitz models, as well as novel
integral representations for the retardation and relaxation spectra in the
Lambert model. In order to derive these analytical expressions, we have used the analytical
properties of the exponential integral and the Lambert $W$ function, as well as the Titchmarsh’s inversion formula of the Stieltjes transform.
In addition, we prove some interesting inequalities comparing
the different models considered, as well as the non-negativity of the
retardation and relaxation spectral functions. This means to prove the
complete monotonicity of the rate of creep and the relaxation functions,
respectively, as required by the classical theory of linear viscoelasticity.
integral; Lambert W function; linear viscoelastic models.
\\
{\bf Publication:}
Mathematics (MDPI), Vol 2024, No 12, 3426/1-14
\\ DOI: 10.3390/math12213426 (Open access)/
\\
{\bf Dedication:}
The authors dedicate this article to their colleague \\ Professor Paolo Emilio RICCI  on the occasion of his 80th birthday.

\end{abstract}



\section{Introduction}

According to the theory of linear viscoelasticity, a material under a unidirectional loading can be
represented as a linear system where either stress, $\sigma \left( t\right) $%
, or strain, $\epsilon \left( t\right) $, serve as input (excitation
function) or output (response function) respectively. Considering that
stress and strain are scaled with respect to suitable reference states, the
input in a creep test is $\sigma \left( t\right) =\theta \left( t\right) $,
while in a relaxation test, it is $\epsilon \left( t\right) =\theta \left(
t\right) $, where $\theta \left( t\right) $ denotes the Heaviside function.
The corresponding outputs are described by time-dependent material
functions. For the creep test, the output is defined as the creep compliance
$J\left( t\right) =\epsilon \left( t\right) $, and for the relaxation test,
the output is defined as the relaxation modulus $G\left( t\right) =\sigma
\left( t\right) $. Experimentally, $J\left( t\right) $ is always
non-decreasing and non-negative, while $G\left( t\right) $ is non-increasing
and non-negative.

According to Gross \cite{Gross BOOK53}, it is quite common to require the existence
of non-negative retardation $R_{\epsilon }\left( \tau \right) $\ and
relaxation $R_{\sigma }\left( \tau \right) $ spectra for the material
functions $J\left( t\right) $ and $G\left( t\right) $, respectively. These
functions are defined as \cite[Eqn. 2.30b]{Mainardi BOOK22}:%
\begin{eqnarray}
J\left( t\right)  &=&a\int_{0}^{\infty }R_{\epsilon }\left( \tau \right)
\,\left( 1-e^{-t/\tau }\right) \,d\tau ,  \label{Re_int_def} \\
G\left( t\right)  &=&b\int_{0}^{\infty }R_{\sigma }\left( \tau \right)
\,e^{-t/\tau }d\tau .  \label{Rs_int_def}
\end{eqnarray}

From a mathematical point of view,  these requirements are equivalent to
state that $J\left( t \right)$ is a Bernstein function and that $G\left( t \right)$ is a completely monotone (CM) function.
We recall that the derivative of a Bernstein function is a CM function, and that any CM
function can be expressed as the Laplace transform of a non-negative
function, that we refer to as the corresponding spectral function or simply
spectrum. For details on Bernstein and CM functions, we refer the reader to
the excellent treatise by Schilling et al. \cite{Schilling-et-al BOOK12}

In order to calculate $R_{\epsilon }\left( \tau \right) $\ and $R_{\sigma
}\left( \tau \right) $ from $J\left( t\right) $ and $G\left( t\right) $,
Gross introduced the frequency spectral functions $S_{\epsilon }\left(
\omega \right) $ and $S_{\sigma }\left( \omega \right) $, defined as 
\cite[Eqn. 2.32]{Mainardi BOOK22}%
\begin{eqnarray}
\,S_{\epsilon }\left( \omega \right)  &=&a\,\frac{R_{\epsilon }\left( \frac{1%
}{\omega }\right) }{\omega ^{2}},  \label{Se_def} \\
\,S_{\sigma }\left( \omega \right)  &=&b\,\frac{R_{\sigma }\left( \frac{1}{%
\omega }\right) }{\omega ^{2}},  \label{Ss_def}
\end{eqnarray}%
where $ \omega =\frac{1}{\tau }$. Therefore, taking the scaling factors $a=b=1
$, we have
\begin{equation}
R_{\epsilon }\left( \tau \right) =\frac{1}{\tau ^{2}}\,S_{\epsilon }\left(
\frac{1}{\tau }\right) ,  \label{R_e(tau)_def}
\end{equation}%
and
\begin{equation}
R_{\sigma }\left( \tau \right) =\frac{1}{\tau ^{2}}\,S_{\sigma }\left( \frac{%
1}{\tau }\right) .  \label{R_sigma(tau)_def}
\end{equation}

In existing literature, many viscoelastic models for the material functions $%
J\left( t\right) $ and $G\left( t\right) $ have been proposed in order to
describe the experimental evidence. The first pioneer to work on linear
viscoelastic models was Richard Becker ($1887-1955$). In $1925$, Becker
introduced a creep law to deal with the deformation of particular
viscoelastic and plastic bodies on the basis of empirical arguments 
\cite{Becker 25}.
This creep law has found applications in ferromagnetism 
\cite{Becker-Doring BOOK39}, and in dielectrics \cite{Gross BOOK53}.
\\
This model has been generalized in \cite{Mainardi-Masina-Spada MTDM19} by using a generalization of the exponential integral based on the Mittag-Leffler function.
In $1956$, Cinna Lomnitz ($1925-2016$) introduced a 
logarithmic creep law to treat the creep behaviour
of igneous rocks \cite{Lomnitz JG56}. This law was also used by Lomnitz to
explain the damping of the free core nutation of the Earth
 \cite{Lomnitz JAP57}, and
the behavior of seismic S-waves \cite{Lomnitz  JGR62}.
In order to generalize the Lomnitz model, 
Harold Jeffreys introduced a new model depending on a parameter $\alpha \in \left[0,1 \right]$ in $1958$ \cite{Jeffreys 58}.
When $\alpha \rightarrow 0$, the Jeffreys model is reduced to the Lomnitz model.
In \cite[Sect. 2.10.1]{Mainardi BOOK22}, the Jeffreys model is extended to negative values of $\alpha$.
Very recently, Mainardi, Masina and Gonzales Santander  have
proposed a new model based on the Lambert $W$ function 
\cite{Mainardi-Masina-Santander SYM23}.
In order to evaluate the principal features of this new model, the main aim of the paper is to compare it with respect to
the Becker and Lomnitz models, since these classical models have several applications and generalizations in the literature.

This paper is organized as follows. In Section \ref{Section: Essentials
viscoelasticity}, we introduce some basic concepts of linear viscoelasticity
that will be used throughout the article. Thereby, we present the
dimensionless relaxation modulus $\phi \left( t\right) $, as well as the
frequency spectral functions $\,S_{\epsilon }\left( \omega \right) $ and $%
S_{\sigma }\left( \omega \right) $ in terms of the Laplace transform.
Sections \ref{Section: Becker model}, \ref{Section:Lomnitz model}, and \ref%
{Section: Lambert model} are devoted to the derivation of analytical
expressions for $\phi \left( t\right) $, as well as $R_{\epsilon }\left(
\tau \right) $ and $R_{\sigma }\left( \tau \right) $ in Becker, Lomnitz, and
Lambert models, respectively. As far as authors' knowledge, the closed-form
expressions found for $R_{\sigma }\left( \tau \right) $ in Becker and
Lomnitz models are novel. Also, the integral expressions found for $%
R_{\epsilon }\left( \tau \right) $ and $R_{\sigma }\left( \tau \right) $ in
Lambert model are also novel. In Section \ref{Section: Numerical results},
we graphically compare the models considered here. Finally, we collect our
conclusions in Section \ref{Section: Conclusions}.

\section{Essentials of linear viscoelasticity \label{Section: Essentials
viscoelasticity}}

In Earth rheology, the law of creep is usually written as:%
\begin{equation}
J\left( t\right) =J_{U}\left[ 1+q\,\psi \left( t\right) \right] ,\quad t\geq
0,  \label{J(t)_def}
\end{equation}%
where $t$ is time, $J_{U}$ is the unrelaxed compliance, $q$ is a positive
dimensionless material constant, and $\psi \left( t\right) $ is the
dimensionless creep function.
Note that the scaling factor $q$ takes into account the effect of different types of materials following the same creep model,
i.e. the same dimensionless creep function $\psi \left( t\right) $.
Since $J\left( 0\right) =J_{U}$, we have
\[
\psi \left( 0\right) =0.
\]

The dimensionless relaxation function is defined as%
\[
\phi \left( t\right) =J_{U}\,G\left( t\right) ,
\]%
where $G\left( t\right) $ is the relaxation modulus. Note that $J_{U}$ acts as a scaling factor for the dimensionless relaxation modulus $\phi \left( t\right)$.
This dimensionless relaxation function obeys the Volterra integral equation \cite[Eqn. 2.89]{Mainardi BOOK22}:
\begin{equation}
\phi \left( t\right) =1-q\int_{0}^{t}\psi ^{\prime }\left( u\right) \,\phi
\left( t-u\right) \,du.  \label{Volterra_phi}
\end{equation}%
From (\ref{Volterra_phi}), we have
\begin{equation}
\phi \left( 0\right) =1.  \label{phi(0)=1}
\end{equation}%
Assuming $q=1$, the solution of (\ref{Volterra_phi})\ is (see Appendix \ref{Appendix: calculation phi})%
\begin{equation}
\phi \left( t\right) =\mathcal{L}^{-1}\left[ \frac{1}{s\left( 1+\mathcal{L}%
\left[ \psi ^{\prime }\left( t\right) ;s\right] \right) };t\right] ,
\label{phi_inverse_Laplace}
\end{equation}%
where $\mathcal{L}\left[ f\left( t\right) ;s\right] =\int_{0}^{\infty
}e^{-st}\,f\left( t\right) \,dt$ denotes the Laplace transform of the
function $f\left( t\right) $ and $\mathcal{L}^{-1}\left[ f\left( t\right) ;s%
\right] $ its inverse Laplace transform.

\subsection{Frequency spectral functions}

From the definitions given in (\ref{Re_int_def})-(\ref{Ss_def}), we can
derive that the creep rate $J^{\prime }\left( t\right) $\ can be expressed
in terms of the frequency spectral function $S_{\epsilon }\left( \omega
\right) $ as (see \cite[Eqn. 2.35]{Mainardi BOOK22}),
\[
J^{\prime }\left( t\right) =\mathcal{L}\left[ \omega \,S_{\epsilon }\left(
\omega \right) ;t\right] ,
\]%
and the relaxation modulus rate $G^{\prime }\left( t\right) $ can be
expressed in terms of the frequency spectral function $S_{\sigma }\left(
\omega \right) $ as
\begin{equation}
G^{\prime }\left( t\right) =-\mathcal{L}\left[ \omega \,S_{\sigma }\left(
\omega \right) ;t\right] .  \label{G'(t)=Laplace}
\end{equation}

On the one hand, taking the scaling factors $q=1$ and $J_{U}=1$, and
differentiating in (\ref{J(t)_def}),\ we have that%
\[
\psi ^{\prime }\left( t\right) =\mathcal{L}\left[ \omega \,S_{\epsilon
}\left( \omega \right) ;t\right] ,
\]%
thus
\[
\mathcal{L}\left[ \psi ^{\prime }\left( t\right) ;s\right] =\mathcal{S}\left[
\omega \,S_{\epsilon }\left( \omega \right) ;s\right] ,
\]%
where $\mathcal{S}\left[ f\left( t\right) ;s\right] =\int_{0}^{\infty
} \frac{f\left( t \right)}{t+s} \,dt $ denotes the Stieltjes
transform \cite[Eqn. 1.14.47]{Olver NIST10}. Applying Titchmarsh's inversion formula \cite[Sect. 3.3]{Apelblat BOOK12}, we obtain%
\begin{equation}
S_{\epsilon }\left( \omega \right) =\frac{1}{\pi \,\omega }\,\mathrm{Im}\,%
\mathcal{L}\left[ \psi ^{\prime }\left( t\right) ;s\right] _{s=\omega
\,e^{-i\pi }}.  \label{S_e(w)_def}
\end{equation}

On the other hand, taking again $J_{U}=1$ in (\ref{J(t)_def}), from (\ref%
{G'(t)=Laplace}), we have
\[
\phi ^{\prime }\left( t\right) =\,G^{\prime }\left( t\right) =-\mathcal{L}%
\left[ \omega \,S_{\sigma }\left( \omega \right) ;t\right] ,
\]%
thus%
\begin{equation}
\mathcal{L}\left[ \phi ^{\prime }\left( t\right) ;s\right] =-\mathcal{S}%
\left[ \omega \,S_{\sigma }\left( \omega \right) ;s\right] .
\label{Laplace[phi']}
\end{equation}%
However,
\[
\mathcal{L}\left[ \phi ^{\prime }\left( t\right) ;s\right] =s\,\mathcal{L}%
\left[ \phi \left( t\right) ;s\right] -\phi \left( 0\right) ,
\]%
thereby, according to (\ref{phi(0)=1}) and (\ref{Laplace[phi']}), we obtain%
\[
1-s\,\mathcal{L}\left[ \phi \left( t\right) ;s\right] =\mathcal{S}\left[
\omega \,S_{\sigma }\left( \omega \right) ;s\right] .
\]%
Apply Titchmarsh's inversion formula again to arrive at%
\begin{eqnarray*}
S_{\sigma }\left( \omega \right) &=&\frac{1}{\pi \,\omega }\,\mathrm{Im}%
\,\left\{ 1-s\,\mathcal{L}\left[ \phi \left( t\right) ;s\right] \right\}
_{s=\omega \,e^{-i\pi }} \\
&=&-\frac{1}{\pi \,\omega }\mathrm{Im}\,\left\{ s\,\mathcal{L}\left[ \phi
\left( t\right) ;s\right] \right\} _{s=\omega \,e^{-i\pi }}.
\end{eqnarray*}%
Finally, according to (\ref{phi_inverse_Laplace}), we have%
\begin{equation}
S_{\sigma }\left( \omega \right) =-\frac{1}{\pi \,\omega }\,\mathrm{Im}%
\,\left\{ \frac{1}{1+\mathcal{L}\left[ \psi ^{\prime }\left( t\right) ;s%
\right] }\right\} _{s=\omega \,e^{-i\pi }}.  \label{S_sigma(w)_def}
\end{equation}

As pointed out in (\ref{R_e(tau)_def}) and (\ref{R_sigma(tau)_def}), the retardation $R_{\epsilon }\left( \tau \right) $\ and
relaxation $R_{\sigma }\left( \tau \right) $ spectra are obtained from the frequency spectral functions
$S_\epsilon(\omega)$ and $S_\sigma(\omega)$ by considering $ \omega =\frac{1}{\tau }$.

\section{Becker model \label{Section: Becker model}}

The law of creep for the Becker model is
\[
\psi _{B}\left( t\right) =\mathrm{Ein}\,\left( \frac{t}{\tau _{0}}\right)
,\quad t\geq 0,\ \tau _{0}>0,
\]%
where the \textit{complementary exponential integral} is defined as
 \cite[Eqn. 6.2.3]{Olver NIST10}%
\[
\mathrm{Ein}\,\left( t\right) =\int_{0}^{t}\frac{1-e^{-u}}{u}\,du,
\]%
being this function an entire function. For simplicity, we take the scaling
factor $\tau _{0}=1$, thus
\begin{equation}
\psi _{B}^{\prime }\left( t\right) =\frac{1-e^{-t}}{t}.
\label{psi'(t)_Becker}
\end{equation}

Taking $\alpha =0$ and $\beta =1$ in the following Laplace transform 
\cite[Eqn. 2.2.4(14)]{Prudnikov-et-al BOOK86}:%
\begin{eqnarray*}
&&\mathcal{L}\left[ \frac{e^{-\alpha t}-e^{-\beta t}}{t};s\right] =\log
\left( \frac{s+\beta }{s+\alpha }\right) , \\
&&\mathrm{Re}\,s>-\mathrm{Re}\,\alpha ,\,\ \mathrm{Re}\,s>-\mathrm{Re}%
\,\beta ,
\end{eqnarray*}%
we conclude that, 
\begin{equation}
\mathcal{L}\left[ \psi _{B}^{\prime }\left( t\right) ;s\right] =\log \left(
1+\frac{1}{s}\right) ,\quad \mathrm{Re}\,s>0,  \label{L[psi']_Becker}
\end{equation}%
Consequently, according to (\ref{phi_inverse_Laplace}) and (\ref%
{L[psi']_Becker}), the dimensionless relaxation modulus is given by
\begin{equation}
\phi _{B}\left( t\right) =\mathcal{L}^{-1}\left[ \frac{1}{s\left( 1+\log
\left( 1+\frac{1}{s}\right) \right) };t\right] .  \label{phi(t)_Becker}
\end{equation}

\subsection{Frequency spectral functions}

Now, according to (\ref{S_e(w)_def}) and (\ref{L[psi']_Becker}), we have
\begin{eqnarray*}
S_{\epsilon }^{B}\left( \omega \right) &=&\frac{1}{\pi \,\omega }\mathrm{Im}%
\,\left[ \log \left( 1-\frac{1}{\omega }\right) \right] \\
&=&\frac{1}{\pi \,\omega }\mathrm{Im}\,\left[ \log \left( \omega -1\right)
-\log \omega \right] .
\end{eqnarray*}%
For $\omega >0$, we have that $\mathrm{Im}\,\left( \log \omega \right) =0$,
hence
\[
S_{\epsilon }^{B}\left( \omega \right) =\frac{1}{\pi \,\omega }\mathrm{Im}\,%
\left[ \log \left\vert \omega -1\right\vert +i\,\mathrm{Arg}\,\left( \omega
-1\right) \right] ,
\]%
where $\mathrm{Arg}\, \left( z \right) \in \left( -\pi, \pi\right] $ denotes the principal argument of $z \in \mathbb{C}$. Therefore, %
\begin{equation}
S_{\epsilon }^{B}\left( \omega \right) =\left\{
\begin{array}{ll}
0, & \omega \geq 1, \\
\displaystyle%
\frac{1}{\,\omega }, & 0\leq \omega <1.%
\end{array}%
\right.  \label{S_e(w)_Becker}
\end{equation}

Also, from (\ref{S_sigma(w)_def}) and (\ref{L[psi']_Becker}), we have%
\begin{eqnarray*}
S_{\sigma }^{B}\left( \omega \right) &=&-\frac{1}{\pi \,\omega }\mathrm{Im}\,%
\left[ \frac{1}{1+\log \left( 1-\frac{1}{\omega }\right) }\right] \\
&=&-\frac{1}{\pi \,\omega }\mathrm{Im}\,\left[ \frac{1}{1+\log \left\vert 1-%
\frac{1}{\omega }\right\vert +i\,\mathrm{Arg}\,\left( 1-\frac{1}{\omega }%
\right) }\right] .
\end{eqnarray*}%
Consider $\omega \geq 0$. Note that $\forall \omega \geq 1$, we have $%
S_{\sigma }^{B}\left( \omega \right) =0$. Also, $\forall \omega \in \left[
0,1\right) $, we have%
\begin{eqnarray*}
S_{\sigma }^{B}\left( \omega \right) &=&-\frac{1}{\pi \,\omega }\mathrm{Im}\,%
\left[ \frac{1}{1+\log \left\vert 1-\frac{1}{\omega }\right\vert +i\,\pi }%
\right] \\
&=&\frac{1}{\omega \left[ \left( 1+\log \left\vert 1-\frac{1}{\omega }%
\right\vert \right) ^{2}+\,\pi ^{2}\right] },
\end{eqnarray*}%
so we conclude%
\begin{equation}
S_{\sigma }^{B}\left( \omega \right) =\left\{
\begin{array}{ll}
0, & \omega \geq 1, \\
\displaystyle%
\frac{1}{\omega \left[ \left( 1+\log \left\vert 1-\frac{1}{\omega }%
\right\vert \right) ^{2}+\,\pi ^{2}\right] }, & 0\leq \omega <1.%
\end{array}%
\right.  \label{S_sigma(w)_Becker}
\end{equation}

\subsection{Retardation and relaxation spectra}

Finally, applying (\ref{R_e(tau)_def}), and taking into account (\ref%
{S_e(w)_Becker}), we arrive at%
\begin{equation}
0\leq R_{\epsilon }^{B}\left( \tau \right) =\left\{
\begin{array}{ll}
0, & 0\leq \tau \leq 1, \\
\displaystyle%
\frac{1}{\tau }, & \tau >1,%
\end{array}%
\right.  \label{ReBecker(t)}
\end{equation}%
which is given in \cite[Eqn. 2.91]{Mainardi BOOK22}. 
Similarly, applying (\ref{R_sigma(tau)_def}), and taking into account (\ref{S_sigma(w)_Becker}), we
obtain%
\begin{equation}
0\leq R_{\sigma }^{B}\left( \tau \right) =\left\{
\begin{array}{ll}
0, & 0\leq \tau \leq 1, \\
\displaystyle%
\frac{1}{\tau \left[ \left( 1+\log \left\vert 1-\tau \right\vert \right)
^{2}+\,\pi ^{2}\right] }, & \tau >1.%
\end{array}%
\right.  \label{RsBecker(t)}
\end{equation}

\section{Lomnitz model \label{Section:Lomnitz model}}

The law of creep in the Lomnitz model is \cite{Lomnitz JG56}:
\[
\psi _{L}\left( t\right) =\log \left( 1+\frac{t}{\tau _{0}}\right) ,\quad
t\geq 0,\ \tau _{0}>0.
\]%
For simplicity, we take the scaling factor $\tau _{0}=1$, thus
\begin{equation}
\psi _{L}^{\prime }\left( t\right) =\frac{1}{1+t}.  \label{psi'(t)_Lomnitz}
\end{equation}

Let us calculate the Laplace transform of $\psi _{L}^{\prime }\left(
t\right) $. Indeed,
\[
\mathcal{L}\left[ \psi _{L}^{\prime }\left( t\right) ;s\right]
=\int_{0}^{\infty }\frac{e^{-st}}{1+t}dt.
\]%
Perform the change of variables $u=1+t,$ and $z=su$ to obtain%
\[
\mathcal{L}\left[ \psi _{L}^{\prime }\left( t\right) ;s\right]
=e^{s}\int_{1}^{\infty }\frac{e^{-su}}{u}du=e^{s}\int_{s}^{\infty }\frac{%
e^{-z}}{z}dz,
\]%
i.e.
\begin{equation}
\mathcal{L}\left[ \psi _{L}^{\prime }\left( t\right) ;s\right]
=e^{s}E_{1}\left( s\right) ,\quad s\neq 0.  \label{L[psi']_Lomnitz}
\end{equation}%
where $E_{1}\left( s\right) $ denotes the\textit{\ exponential integral}
\cite[Eqn. 6.2.1]{Olver NIST10}. According to (\ref{phi_inverse_Laplace}) and (\ref{L[psi']_Lomnitz}), the dimensionless relaxation modulus is given by
\begin{equation}
\phi _{L}\left( t\right) =\mathcal{L}^{-1}\left[ \frac{1}{s\left(
1+e^{s}E_{1}\left( s\right) \right) };t\right] .  \label{phi(t)_Lomnitz}
\end{equation}

\subsection{Frequency spectral functions}

According to (\ref{S_e(w)_def}) and (\ref{L[psi']_Lomnitz}), we have%
\[
S_{\epsilon }^{L}\left( \omega \right) =\frac{1}{\pi \,\omega }\,\mathrm{Im}%
\,\left[ e^{-\omega }E_{1}\left( \omega \,e^{-i\pi }\right) \right] .
\]%
The exponential integral function $E_{1}\left( z\right) $ can be expressed
as \cite[Eqn. 6.2.2]{Olver NIST10}%
\begin{equation}
E_{1}\left( z\right) =\mathrm{Ein}\,\left( z\right) -\log z-\gamma ,
\label{E1_prop}
\end{equation}%
where $\gamma $ denotes the Euler-Mascheroni constant. Since the power
series of the complementary exponential integral is given by 
\cite[Eqn. 6.6.4]{Olver NIST10}%
\[
\mathrm{Ein}\,\left( z\right) =-\sum_{k=1}^{\infty }\frac{\left( -z\right)
^{k}}{k!\,k},\quad \left\vert z\right\vert <\infty ,
\]%
it is apparent that
\begin{equation}
\forall x\in
\mathbb{R}
,\ \mathrm{Ein}\,\left( x\right) \in
\mathbb{R}
.  \label{Ein(z)_real}
\end{equation}%
Therefore, $\forall \omega >0$, we have%
\begin{eqnarray*}
S_{\epsilon }^{L}\left( \omega \right) &=&\frac{e^{-\omega }}{\pi \,\omega }%
\,\mathrm{Im}\,\left[ E_{1}\left( \omega \,e^{-i\pi }\right) \right] \\
&=&-\frac{e^{-\omega }}{\pi \,\omega }\,\mathrm{Im}\,\left[ \log \left(
\omega \,e^{-i\pi }\right) \right] ,
\end{eqnarray*}%
so that%
\begin{equation}
S_{\epsilon }^{L}\left( \omega \right) =\frac{e^{-\omega }}{\omega },\quad
\omega >0.  \label{S_e(w)_Lomnitz}
\end{equation}

Also, from (\ref{S_sigma(w)_def}) and (\ref{L[psi']_Lomnitz}), we have%
\[
S_{\sigma }^{L}\left( \omega \right) =-\frac{1}{\pi \,\omega }\mathrm{Im}\,%
\left[ \frac{1}{1+e^{-\omega }E_{1}\left( \omega \,e^{-i\pi }\right) }\right]
.
\]%
For $\omega >0$, and taking into account (\ref{E1_prop}),
\begin{eqnarray*}
S_{\sigma }^{L}\left( \omega \right) &=&-\frac{e^{\omega }}{\pi \,\omega }%
\mathrm{Im}\,\left[ \frac{1}{e^{\omega }+E_{1}\left( \omega \,e^{-i\pi
}\right) }\right] \\
&=&-\frac{e^{\omega }}{\pi \,\omega }\mathrm{Im}\,\left[ \frac{1}{e^{\omega
}+\mathrm{Ein}\,\left( -\omega \right) -\log \left( \omega \,e^{-i\pi
}\right) -\gamma }\right] \\
&=&-\frac{e^{\omega }}{\pi \,\omega }\mathrm{Im}\,\left[ \frac{1}{e^{\omega
}+\mathrm{Ein}\,\left( -\omega \right) -\log \omega \,-\gamma +i\,\pi }%
\right] \\
&=&-\frac{e^{\omega }}{\pi \,\omega }\mathrm{Im}\,\left[ \frac{e^{\omega }+%
\mathrm{Ein}\,\left( -\omega \right) -\log \omega \,-\gamma -i\,\pi }{\left(
e^{\omega }+\mathrm{Ein}\,\left( -\omega \right) -\log \omega \,-\gamma
\right) ^{2}+\pi ^{2}}\right] ,
\end{eqnarray*}%
i.e.%
\begin{equation}
S_{\sigma }^{L}\left( \omega \right) =\frac{e^{\omega }}{\omega \left[
\left( e^{\omega }+\mathrm{Ein}\,\left( -\omega \right) -\log \omega
\,-\gamma \right) ^{2}+\pi ^{2}\right] },\quad \omega >0.
\label{S_sigma(w)_Lomnitz}
\end{equation}

\subsection{Retardation and relaxation spectra}

Finally, applying (\ref{R_e(tau)_def}), and taking into account (\ref%
{S_e(w)_Lomnitz}), we conclude%
\begin{equation}
R_{\epsilon }^{L}\left( \tau \right) =\frac{e^{-1/\tau }}{\tau }>0,\quad
\tau >0,  \label{ReLomnitz(t)}
\end{equation}%
which is given in \cite[Eqn. 2.92]{Mainardi BOOK22}.
Similarly, applying (\ref%
{R_sigma(tau)_def}), and taking into account (\ref{S_sigma(w)_Becker}), we
conclude%
\begin{equation}
R_{\sigma }^{L}\left( \tau \right) =\frac{e^{1/\tau }}{\tau \left[ \left(
e^{1/\tau }+\mathrm{Ein}\,\left( -\frac{1}{\tau }\right) +\log \tau
\,-\gamma \right) ^{2}+\pi ^{2}\right] }>0,\quad \tau >0.
\label{RsLomnitz(t)}
\end{equation}

\section{Lambert model \label{Section: Lambert model}}

The law of creep in the Lambert model is 
\cite{Mainardi-Masina-Santander SYM23}:
\[
\psi _{W}\left( t\right) =W_{0}\left( t\right) ,\quad t\geq 0,
\]%
where $W_{0}\left( t\right) $ denotes the principal branch of the Lambert $W$
function \cite[Sect. 4.13]{Olver NIST10}. The Lambert $W$ function $W\left( z\right)
$ is defined as the root of the transcendental equation:\
\begin{equation}
W\left( z\right) \exp \left( W\left( z\right) \right) =z.  \label{W_def}
\end{equation}%
For $z\geq 0$, (\ref{W_def})\ has only one real solution, which is denoted
as $W_{0}\left( z\right) $. Therefore, from (\ref{W_def}), we have%
\begin{equation}
\psi _{W}\left( 0\right) =W_{0}\left( 0\right) =0.  \label{psi_W(0)=0}
\end{equation}%
Differentiating in (\ref{W_def}), it is easy to prove that
\begin{equation}
\psi _{W}^{\prime }\left( t\right) =W_{0}^{\prime }\left( t\right) =\frac{1}{%
\exp \left( W_{0}\left( t\right) \right) +t}=\frac{W_{0}\left( t\right) }{%
t\left( 1+W_{0}\left( t\right) \right) }.  \label{psi(t)_Lambert}
\end{equation}

Taking into account (\ref{psi_W(0)=0}), the Laplace transform of the creep
rate is,%
\[
\mathcal{L}\left[ \psi _{W}^{\prime }\left( t\right) ;s\right] =s\,\mathcal{L%
}\left[ \psi _{W}\left( t\right) ;s\right] -\psi _{W}\left( 0\right) =s\,%
\mathcal{L}\left[ W_{0}\left( t\right) ;s\right] ,
\]%
thus, according to (\ref{phi_inverse_Laplace}), we have for the Lambert model,
\begin{equation}
\phi _{W}\left( t\right) =\mathcal{L}^{-1}\left[ \frac{1}{s\left( 1+s\,%
\mathcal{L}\left[ W_{0}\left( t\right) ;s\right] \right) };t\right] ,
\label{phi(t)_Lambert}
\end{equation}%

\subsection{Retardation and relaxation spectra}
Unfortunately, it is not known an analytical expression for the Laplace
transform of the Lambert $W$ function, so that the dimensionless relaxation
modulus $\phi _{W}\left( t\right) $ has to be numerically evaluated, as well
as the frequency spectral functions:\
\begin{eqnarray}
S_{\epsilon }^{W}\left( \omega \right) &=&\frac{1}{\pi \,\omega }\mathrm{Im}%
\left\{ \,s\,\mathcal{L}\left[ W_{0}\left( t\right) ;s\right] \right\}
_{s=\omega \,e^{-i\pi }},  \label{SeLambert_a} \\
S_{\sigma }^{W}\left( \omega \right) &=&-\frac{1}{\pi \,\omega }\mathrm{Im}\,%
\left[ \frac{1}{1+s\,\mathcal{L}\left[ W_{0}\left( t\right) ;s\right] }%
\right] _{s=\omega \,e^{-i\pi }},  \label{SsLambert_a}
\end{eqnarray}%
from which we obtain  the retardation and the relaxation spectra, respectively,%
\begin{eqnarray*}
R_{\epsilon }^{W}\left( \tau \right) &=&\frac{1}{\tau ^{2}}\,S_{\epsilon
}^{W}\left( \frac{1}{\tau }\right) , \\
R_{\sigma }^{W}\left( \tau \right) &=&\frac{1}{\tau ^{2}}\,S_{\sigma
}^{W}\left( \frac{1}{\tau }\right) .
\end{eqnarray*}

For computational purposes, we are going to reduce (\ref{SeLambert_a}) and (%
\ref{SsLambert_a}) as follows:\
\[
S_{\epsilon }^{W}\left( \omega \right) =\frac{1}{\pi \,\omega }\mathrm{Im}%
\left[ \,s\,\int_{0}^{\infty }e^{-st}W_{0}\left( t\right) dt\right]
_{s=\omega \,e^{-i\pi }}.
\]%
Perform the change of variables $z=st$, to obtain%
\[
S_{\epsilon }^{W}\left( \omega \right) =\frac{1}{\pi \,\omega }\mathrm{Im}%
\left[ \int_{0}^{\infty }e^{-z}W_{0}\left( \frac{z}{s}\right) dz\right]
_{s=\omega \,e^{-i\pi }}.
\]%
For $\omega >0$, we have%
\[
S_{\epsilon }^{W}\left( \omega \right) =\frac{1}{\pi \,\omega }%
\int_{0}^{\infty }e^{-z}\,\mathrm{Im\,}\left[ W_{0}\left( -\frac{z}{\omega }%
\right) \right] \,dz,
\]%
thus%
\begin{equation}
R_{\epsilon }^{W}\left( \tau \right) =\frac{1}{\pi \,\tau }%
\,\int_{0}^{\infty }e^{-z}\,\mathrm{Im\,}\left[ W_{0}\left( -z\,\tau \right) %
\right] \,dz,\quad \tau \geq 0.  \label{ReLambert(t)}
\end{equation}%
According to the following bound (see Appendix \ref{Appendix: imaginary Lambert}),%
\begin{equation}
0\leq \mathrm{Im\,}W_{0}\left( x\right) <\pi ,\quad x<0,
\label{Bound_Lambert}
\end{equation}%
it is apparent that
\begin{equation}
R_{\epsilon }^{W}\left( \tau \right) \geq 0,\quad \tau \geq 0.
\label{Re_Lambert_pos}
\end{equation}%
In addition, if we take into account (\ref{ReBecker(t)}), we conclude
\begin{equation}
R_{\epsilon }^{W}\left( \tau \right) <\frac{1}{\tau }=\,R_{\epsilon
}^{B}\left( \tau \right) ,\quad \tau >1.  \label{ReLambert<ReBecker}
\end{equation}

Similarly, we obtain%
\begin{eqnarray*}
S_{\sigma }^{W}\left( \omega \right) &=&-\frac{1}{\pi \,\omega }\,\mathrm{Im}%
\,\left[ \frac{1}{1+s\,\mathcal{L}\left[ W_{0}\left( t\right) ;s\right] }%
\right] _{s=\omega \,e^{-i\pi }} \\
&=&-\frac{1}{\pi \,\omega }\,\mathrm{Im}\,\left[ \frac{1}{1+\int_{0}^{\infty
}e^{-z}W_{0}\left( \frac{z}{s}\right) dz}\right] _{s=\omega \,e^{-i\pi }}.
\end{eqnarray*}%
For $\omega >0$, we have%
\[
S_{\sigma }^{W}\left( \omega \right) =-\frac{1}{\pi \,\omega }\,\mathrm{Im}\,%
\left[ \frac{1}{1+\int_{0}^{\infty }e^{-z}W_{0}\left( -\frac{z}{\omega }%
\right) dz}\right] ,
\]%
thus%
\begin{equation}
R_{\sigma }^{W}\left( \tau \right) =-\frac{1}{\pi \,\tau }\,\mathrm{Im}\,%
\left[ \frac{1}{1+\int_{0}^{\infty }e^{-z}W_{0}\left( -z\,\tau \right) dz}%
\right] ,\quad \tau \geq 0.  \label{RsLambert(t)}
\end{equation}

According to the following property,
\[
\forall w\in
\mathbb{C}
,\quad \mathrm{Im}\,\left[ \frac{1}{w}\right] =-\frac{\mathrm{Im}\,\left[ w%
\right] }{\left\vert w\right\vert ^{2}},
\]%
we rewrite (\ref{RsLambert(t)})\ as
\[
R_{\sigma }^{W}\left( \tau \right) =\frac{\int_{0}^{\infty }e^{-z}\,\mathrm{%
Im\,}\left[ W_{0}\left( -z\,\tau \right) \right] \,dz}{\pi \,\tau \left\vert
1+\int_{0}^{\infty }e^{-z}W_{0}\left( -z\,\tau \right) dz\right\vert ^{2}}%
\,,\quad \tau \geq 0,
\]%
thus, taking into account the bound given in (\ref{Bound_Lambert}), we
conclude that
\begin{equation}
R_{\sigma }^{W}\left( \tau \right) \geq 0,\quad \tau \geq 0.
\label{RsLambert_pos}
\end{equation}

\section{Numerical results \label{Section: Numerical results}}

\begin{figure}[htbp]
\centering\includegraphics[width=8cm]{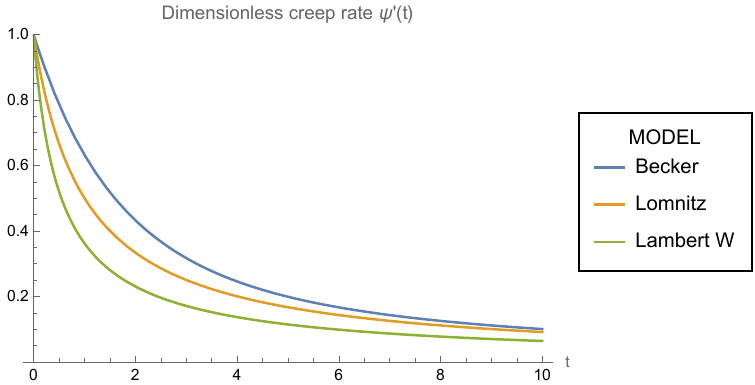}
\caption{Dimensionless creep rate.}
\label{Figure: creep_rate}
\end{figure}

Fig. \ref{Figure: creep_rate} shows the dimensionless creep rate $\psi
^{\prime }\left( t\right) $\ for the Becker (\ref{psi'(t)_Becker}), Lomnitz (\ref%
{psi'(t)_Lomnitz}), and Lambert (\ref{psi(t)_Lambert}) models. It is apparent
that for $t\geq 0$, $\psi ^{\prime }\left( t\right) $ is a monotonic
decreasing function in all models, with $\psi ^{\prime }\left( 0\right) =1$.
Also, Fig. \ref{Figure: creep_rate} suggests that $\psi _{W}^{\prime }\left(
t\right) <\psi _{L}^{\prime }\left( t\right) <\psi _{B}^{\prime }\left(
t\right) $ for $t>0$. The latter can be derived very easily by analytical
methods, as it is shown in Appendix \ref{Appendix: inequality}.

\begin{figure}[htbp]
\centering\includegraphics[width=8cm]{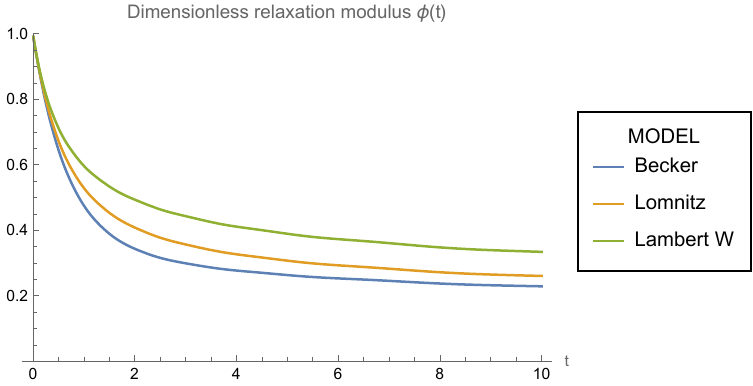}
\caption{Dimensionless relaxation modulus.}
\label{Figure: relaxation_modulus}
\end{figure}

In Fig. \ref{Figure: relaxation_modulus}, the dimensionless relaxation
modulus $\phi \left( t\right) $ is plotted for all the models considered
here. For this purpose, we have numerically evaluated (\ref{phi(t)_Becker}),
(\ref{phi(t)_Lomnitz})\ and (\ref{phi(t)_Lambert}), computing the inverse Laplace transform
with the Papoulis method \cite{Papoulis QAM57} integrated into MATHEMATICA. It is apparent that $\phi \left(
t\right) $ is a monotonic decreasing function for $t\geq 0$ with $\phi
\left( 0\right) =1$ (\ref{phi(0)=1}). Also, Fig. \ref{Figure:
relaxation_modulus} suggests that $\phi _{B}\left( t\right) <\phi _{L}\left(
t\right) <\phi _{W}\left( t\right) $ for $t>0$. Unlike the case of the creep
rate, it does not seem to be any analytical method to derive the latter
inequality.

\begin{figure}[htbp]
\centering\includegraphics[width=8cm]{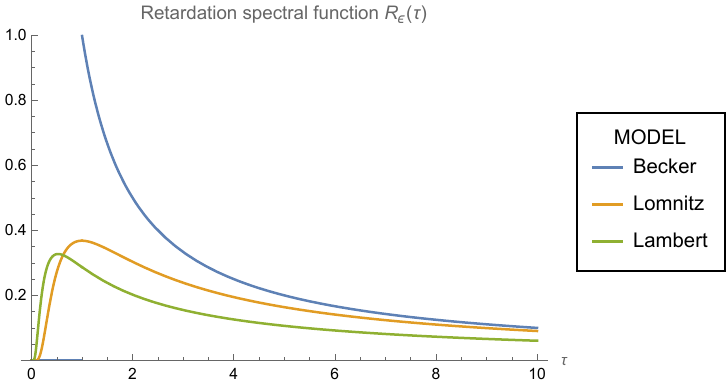}
\caption{Retardation spectral function $R_{\protect\epsilon }\left( \protect%
\tau \right) $.}
\label{Figure: Re(t)}
\end{figure}

Fig. \ref{Figure: Re(t)} presents the retardation spectral function $%
R_{\epsilon }\left( \tau \right) $ for the Becker (\ref{ReBecker(t)}), Lomnitz (%
\ref{ReLomnitz(t)}), and Lambert (\ref{ReLambert(t)})\ models. It is worth
noting that the graphs corresponding to the Lambert and Becker models
verify the inequality derived in (\ref{ReLambert<ReBecker}), i.e. $%
R_{\epsilon }^{W}\left( \tau \right) <\,R_{\epsilon }^{B}\left( \tau \right)
$ for $\tau >1$.

\begin{figure}[htbp]
\centering\includegraphics[width=8cm]{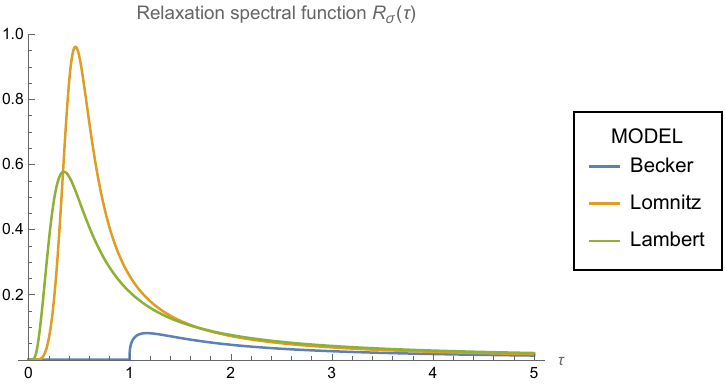}
\caption{Relaxation spectral function $R_{\protect\sigma }\left( \protect%
\tau \right) $.}
\label{Figure: Rs(t)}
\end{figure}

In Fig. \ref{Figure: Rs(t)}, the relaxation spectral function $R_{\sigma
}\left( \tau \right) $ is plotted for the Becker (\ref{RsBecker(t)}), Lomnitz (%
\ref{RsLomnitz(t)}), and Lambert (\ref{RsLambert(t)})\ models.
Despite the fact the plots for $\psi ^{\prime }\left( t\right) $ and $\phi \left(
t\right) $ are quite similar for all models, there is a qualitative great
difference between the Becker model, and the Lomnitz and Lambert models with
respect the spectral functions $R_{\epsilon }\left( \tau \right) $ and $%
R_{\sigma }\left( \tau \right) $, since the Becker model shows a
discontinuity at $\tau =1$, but we have continuous functions for the other
two models.

Finally, according to Figs. \ref{Figure: Re(t)} and \ref{Figure: Rs(t)}, it
is apparent that the retardation $R_{\epsilon }\left( \tau \right) $\ and
relaxation $R_{\sigma }\left( \tau \right) $ spectra for all the models
considered here are non-negative functions, as aforementioned in the
Introduction. This is consistent with what was said above in (\ref%
{ReBecker(t)})-(\ref{RsBecker(t)}) for the Becker model, (\ref{ReLomnitz(t)})-(%
\ref{RsLomnitz(t)})\ for the Lomnitz model, and (\ref{Re_Lambert_pos}), (\ref%
{RsLambert_pos})\ for the Lambert model.

\section{Conclusions \label{Section: Conclusions}}

We have compared some of the viscoelastic models for the law of creep
reported in the literature, i.e. the Becker, Lomnitz and Lambert models. For
these models, we have computed in a much more efficient way the dimensionless
relaxation modulus $\phi \left( t\right) $ by numerically evaluating the
inverse Laplace transform that appears in (\ref{phi_inverse_Laplace}),
instead of numerically solving the Volterra integral equation given in (\ref%
{Volterra_phi}), (see \cite[Sect. 2.9.2]{Mainardi BOOK22}).
The numerical inversion of the Laplace transform has been carried out by using Papoulis method \cite{Papoulis QAM57}.

For the Becker and Lomnitz models, we have given novel closed-form formulas
for the relaxation spectrum $R_{\sigma }\left( \tau \right) $, i.e. Eqns. (%
\ref{RsBecker(t)}), and (\ref{RsLomnitz(t)}).
It is worth noting that these closed-form formulas allow us to compute $R_{\sigma }\left( \tau \right) $ in a much more efficient way than the numerical inversion of (\ref{Rs_int_def}). As aforementioned in the Introduction, the Becker and Lomnitz models have been considered in many physical applications, thus these closed-form formulas are quite valuable in the field of viscoelastic materials.
\newpage
Also as a novelty, we have
derived the retardation and relaxation spectra $R_{\epsilon }\left( \tau \right) $ and $%
R_{\sigma }\left( \tau \right) $ in integral form for the Lambert model in (%
\ref{ReLambert(t)})\ and (\ref{RsLambert(t)}). Again, these integral representations allow us to compute $R_{\epsilon }\left( \tau \right) $ and $%
R_{\sigma }\left( \tau \right) $ very efficiently.

Further, we have proved that
the spectral functions $R_{\epsilon }\left( \tau \right) $ and $R_{\sigma
}\left( \tau \right) $ are non-negative functions for all the models
considered. It is worth noting as well that we have proved the interesting
inequality $R_{\epsilon }^{W}\left( \tau \right) <\,R_{\epsilon }^{B}\left(
\tau \right) $ for $\tau >0$, comparing the retardation spectrum of the
Lambert and Becker models.

In addition, we have compared all the models in terms of the behaviour of
the dimensionless creep rate $\psi ^{\prime }\left( t\right) $ in Fig. \ref%
{Figure: creep_rate}, as well as the dimensionless relaxation modulus $\phi
\left( t\right) $ in Fig. \ref{Figure: relaxation_modulus}, and the spectral functions $R_{\epsilon }\left( \tau \right) $ and $%
R_{\sigma }\left( \tau \right) $ in Figs. \ref{Figure: Re(t)}\ and \ref%
{Figure: Rs(t)}. From these graphs, we can appreciate that all the models
are asymptotically equivalent, although not at the same rate. It is worth
noting that the Becker model presents a discontinuity in the retardation
spectral functions $R_{\epsilon }\left( \tau \right) $ and $R_{\sigma
}\left( \tau \right) $ at $\tau =1$, which is not present in the other
models.

Finally, in Appendix C, we have proved the following inequality for the
creep rate:\
\[
\psi _{W}^{\prime }\left( t\right) <\psi _{L}^{\prime }\left( t\right) <\psi
_{B}^{\prime }\left( t\right) ,\quad t>0.
\]
However, the following inequality for the dimensionless relaxation modulus:
\[
\phi _{B}\left( t\right) <\phi _{L}\left( t\right) <\phi _{W}\left( t\right)
,\quad t>0,
\]%
which is suggested by Fig. \ref{Figure: relaxation_modulus}, seems to be
true, but we could not find any mathematical proof for it.

\section*{Acknowledgments}
 We  thank the anonymous reviewers for their constructive comments and suggestions, which have helped to improve this article.

\section*{Appendix A: Calculation of $\protect\phi \left( t\right) $} \label{Appendix: calculation phi}

Let us solve (\ref{Volterra_phi}), taking $q=1$, i.e.%
\begin{equation}
\int_{0}^{t}\psi ^{\prime }\left( u\right) \,\phi \left( t-u\right)
\,du=1-\phi \left( t\right) .  \label{Eq_Volterra}
\end{equation}%
In order to calculate (\ref{Eq_Volterra}), let us introduce the convolution
theorem for the Laplace transform \cite[Theorem 2.39]{Schiff BOOK13}.
\\
{\bf Theorem} (convolution theorem)
{\it If $f$ and $g$ are piecewise continuous on $\left[ 0,\infty \right) $ and of
exponential order $\alpha $, then%
\[
\mathcal{L}\left[ \left( f\ast g\right) \left( t\right) ;s\right] =\mathcal{L%
}\left[ f\left( t\right) ;s\right] \,\mathcal{L}\left[ g\left( t\right) ;s%
\right] ,\qquad \mathrm{Re}\,s>\alpha ,
\]%
where the convolution is given by the integral%
\[
\left( f\ast g\right) \left( t\right) =\int_{0}^{t}f\left( u\right)
\,g\left( t-u\right) \,du.
\]
} 

Therefore, applying the Laplace transform to (\ref{Eq_Volterra}), we obtain%
\begin{equation}
\mathcal{L}\left[ \psi ^{\prime }\left( t\right) ;s\right] \,\mathcal{L}%
\left[ \phi \left( t\right) ;s\right] =\mathcal{L}\left[ 1-\phi \left(
t\right) ;s\right] .  \label{Laplace_Eq_Volterra}
\end{equation}%
Recall that
\[
\mathcal{L}\left[ 1 ;s \right] =\int_{0}^{\infty }e^{-st}dt=\frac{1}{s},
\]%
in order to rewrite (\ref{Laplace_Eq_Volterra})\ as%
\[
\mathcal{L}\left[ \psi ^{\prime }\left( t\right) ;s\right] \,\mathcal{L}%
\left[ \phi \left( t\right) ;s\right] =\frac{1}{s}-\mathcal{L}\left[ \phi
\left( t\right) ;s\right] .
\]%
Solving for $\phi \left( t\right) $, we conclude that%
\[
\phi \left( t\right) =\mathcal{L}^{-1}\left[ \frac{1}{s\left( 1+\mathcal{L}%
\left[ \psi ^{\prime }\left( t\right) ;s\right] \right) };t \right] .
\]

\section*{Appendix B: The imaginary part of the Lambert function for negative values} \label{Appendix: imaginary Lambert}

We want to bound the imaginary part of the principal branch of the Lambert $%
W $ function for negative values of the argument, i.e. $\mathrm{Im\,}%
W_{0}\left( x\right) $ for $x<0$. Consider $w=W_{0}\left( z\right) $, so
that $z=w\,e^{w}$. If $z=x+i\,y$, and $w=\xi +i\,\eta $, we have that%
\begin{eqnarray*}
x+i\,y &=&\left( \xi +i\,\eta \right) \exp \left( \xi +i\,\eta \right) \\
&=&e^{\xi }\left( \xi +i\,\eta \right) \left( \cos \eta +i\sin \eta \right) ,
\end{eqnarray*}%
thus%
\begin{eqnarray}
x &=&e^{\xi }\left( \xi \cos \eta -\eta \sin \eta \right) ,  \label{x_def} \\
y &=&e^{\xi }\left( \eta \cos \eta +\xi \sin \eta \right) .  \label{y_def}
\end{eqnarray}%
If $z\in
\mathbb{R}
$, i.e. $y=0$, we have $\eta =0$ or $\xi =-\eta \cot \eta $. The case $\eta
=0$ is not interesting for our purpose, since $\eta =\mathrm{Im\,}%
W_{0}\left( z\right) $. Therefore, insert $\xi =-\eta \cot \eta $ in (\ref%
{x_def})\ to obtain%
\[
f\left( \eta \right) :=-\eta \exp \left( -\eta \cot \eta \right) \csc \eta
=x.
\]%
Note that $f\left( \eta \right) $ is an even function. Moreover,
\[
\lim_{\eta \rightarrow 0}f\left( \eta \right) =\lim_{\eta \rightarrow
0}\left( -\frac{\eta }{\sin \eta }\right) \exp \left( -\frac{\eta }{\sin
\eta }\cos \eta \right) =-\frac{1}{e},
\]%
and
\[
\lim_{\eta \rightarrow \pi ^{-}}f\left( \eta \right) =-\infty .
\]

\begin{figure}[tbph]
\centering\includegraphics[width=8cm]{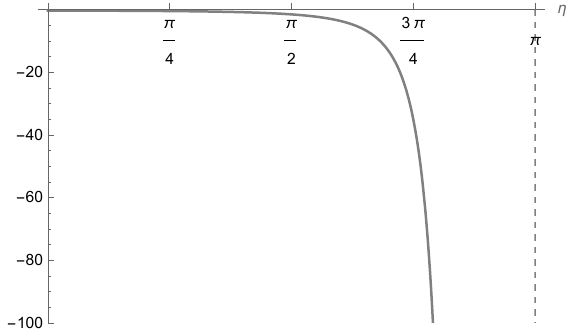}
\caption{Function $f\left( \protect\eta \right) $.}
\label{Figure: f-eta}
\end{figure}
Since Fig. \ref{Figure: f-eta} shows that $f\left( \eta \right) $ is a
monotonic decreasing function for $\eta \in \left( 0,\pi \right) $, we
obtain the following bound,
\[
0<\mathrm{Im\,}W_{0}\left( x\right) <\pi ,\quad x<-\frac{1}{e}.
\]%
Moreover, since $W_{0}\left( x\right) \in
\mathbb{R}
$ for $x\geq -\frac{1}{e}$ \cite[Sect. 4.13]{Olver NIST10}, we conclude%
\begin{equation}
0\leq \mathrm{Im\,}W_{0}\left( x\right) <\pi ,\quad x<0.  \label{Bound_Im_W}
\end{equation}

\section*{Appendix C: The creep rate inequality} \label{Appendix: inequality}

According to (\ref{psi'(t)_Becker}), (\ref{psi'(t)_Lomnitz})\ and (\ref%
{psi(t)_Lambert}), the dimensionless creep rates for the models considered
are%
\begin{eqnarray*}
\psi _{B}^{\prime }\left( t\right) &=&\frac{1-e^{-t}}{t}, \\
\psi _{L}^{\prime }\left( t\right) &=&\frac{1}{1+t}, \\
\psi _{W}^{\prime }\left( t\right) &=&\frac{1}{\exp \left( W_{0}\left(
t\right) \right) +t}.
\end{eqnarray*}

On the one hand, according to (\ref{psi_W(0)=0}), $W_{0}\left( 0\right) =0$,
and according to (\ref{psi(t)_Lambert}), $W_{0}^{\prime }\left( t\right) =%
\frac{1}{\exp \left( W_{0}\left( t\right) \right) +t}>0$ for $t>0$;\ thus $%
W_{0}\left( t\right) >0$, i.e. $\exp \left( W_{0}\left( t\right) \right) >1$%
, for $t>0$. Thereby,
\begin{equation}
\psi _{W}^{\prime }\left( t\right) <\psi _{L}^{\prime }\left( t\right)
,\quad t>0.  \label{psi'_W<psi'L}
\end{equation}

On the other hand, consider $f\left( t\right) =\log \left( 1+t\right) $ and $%
g\left( t\right) =t$. Since $f\left( 0\right) =g\left( 0\right) =0$ and $%
f\left( t\right) $ is a convex function, we have that%
\begin{equation}
t=g\left( t\right) >f\left( t\right) =\log \left( 1+t\right) ,\quad t>0.
\label{ineq_log}
\end{equation}%
From (\ref{ineq_log}), we have for $t>0$%
\begin{eqnarray*}
t>\log \left( 1+t\right) &\leftrightarrow &-t<-\log \left( 1+t\right) =\log
\left( \frac{1}{1+t}\right) \\
&\leftrightarrow &e^{-t}<\frac{1}{1+t} \\
&\leftrightarrow &1-e^{-t}>1-\frac{1}{1+t}=\frac{t}{1+t} \\
&\leftrightarrow &\frac{1-e^{-t}}{t}>\frac{1}{1+t}.
\end{eqnarray*}%
Consequently,
\begin{equation}
\psi _{L}^{\prime }\left( t\right) <\psi _{B}^{\prime }\left( t\right)
,\quad t>0.  \label{psi'L<psi'W}
\end{equation}

In summary, from (\ref{psi'_W<psi'L})\ and (\ref{psi'L<psi'W}), we conclude%
\[
\psi _{W}^{\prime }\left( t\right) <\psi _{L}^{\prime }\left( t\right) <\psi
_{B}^{\prime }\left( t\right) ,\quad t>0.
\]


\end{document}